\documentclass[iop]{emulateapj}

\newcommand{\icarus}{Icarus}

\shorttitle{The distinct spectrum of large-scale chaos}
\shortauthors{Fischer et al.}

\begin{document}

\submitted{Accepted to The Astronomical Journal}

\title{Spatially resolved spectroscopy of Europa: \\
the distinct spectrum of large-scale chaos}

\author{
P. D. Fischer\altaffilmark{1}, 
M. E. Brown\altaffilmark{2}, 
and K. P. Hand\altaffilmark{3}}
\affil{Division of Geological and Planetary Sciences, California Institute of Technology, Pasadena, CA 91125, USA \\
Jet Propulsion Laboratory, California Institute of Technology, Pasadena, CA 91109, USA}
\email{pfischer@caltech.edu}

\begin{abstract}

We present a comprehensive analysis of spatially resolved moderate spectral resolution near infrared spectra obtained with the adaptive optics system at the Keck Observatory. We identify three compositionally distinct end member regions: the trailing hemisphere bullseye, the leading hemisphere upper latitudes, and a third component associated with leading hemisphere chaos units. We interpret the composition of the three end member regions to be dominated by irradiation products, water ice, and evaporite deposits or salt brines, respectively. The third component is associated with geological features and distinct from the geography of irradiation, suggesting an endogenous identity. Identifying the endogenous composition is of particular interest for revealing the subsurface composition. However, its spectrum is not consistent with linear mixtures of the salt minerals previously considered relevant to Europa. The spectrum of this component is distinguished by distorted hydration features rather than distinct spectral features, indicating hydrated minerals but making unique identification difficult.  In particular, it lacks features common to hydrated sulfate minerals, challenging the traditional view of an endogenous salty component dominated by Mg-sulfates. Chloride evaporite deposits are one possible alternative. 
\end{abstract}


\section{Introduction}

Europa's icy, geologically active surface conceals a global, subsurface ocean \citep{kiv00} in contact with a rocky sea floor and potentially hospitable to chemosynthetic life. The greatest challenge for generating and sustaining such a biosphere may be the compositional environment and chemical budget of the subsurface ocean \citep{kph09}. Therefore, we must understand the environment of the subsurface ocean in order to understand its potential for habitability. Surface composition is currently the best measurable link to the subsurface. 

Europa's surface composition is dominated by water ice \citep{kui57}, but contains a variety of non-water-ice species delivered by impacts, emplaced endogenously by geophysical processes, and produced exogenously by irradiation of the pre-existing surface \citep[][and references therein]{car09}. Many of these species are known or speculated to be present. Material ejected from Io is ionized and swept up by the Jovian magnetosphere and implanted onto Europa's surface, particularly ionized sulfur and oxygen. Energetic ions and electrons are also showered onto the surface, as well as cosmic rays and solar radiation, which break apart existing molecular bonds and form new radiolytic species, such as SO$_2$ \citep{lane81}, O$_2$ \citep{lan78, joh82}, H$_2$O$_2$, \citep{car99a}, and H$_2$SO$_4$ and elemental sulfur S$_x$ \citep{car99}. Endogenous material is revealed by association with geologic features; irradiated NaCl \citep{kph15} or sulfur polymers \citep{car99} may cause the red-brown color associated with linea. Na and K have been detected in the sputtered atmosphere, and their abundance ratio may indicate an endogenous source, at least for Na \citep{brown01}. Epsomite has been detected \citep{brown13}, and other sulfate salts are speculated to be present as well \citep{mcc98}, though an exogenous vs. endogenous origin is uncertain \citep{brown13, kph15}. 

Surface composition is one boundary in a continuous cycle of exchange with the subsurface. Upward transport through the ice shell by processes such as compositional diapirism \citep{pap04} or convective ice plumes \citep{ped15} may leave a compositional fingerprint on the surface from the ocean. Downward transport of radiolytic species from the surface, aided by tectonic subsumption \citep{kat14}, may provide the critical fuel for a redox-powered ecosystem \citep{chyba00, kph09} if there is interaction with the ocean in appropriate time scales. Identifying the non-water-ice components is key to understanding the subsurface environment. 

Previous efforts to identify the non-ice surface components and determine their geographical associations is mostly based on measurements from the \textit{Galileo} Near Infrared Mapping Spectrometer (NIMS). \cite{mcc98} recognized the presence of hydrated impurities, and suggested hydrated salts and evaporites to be the non-ice component across the trailing hemisphere and in the linea. \cite{car99} proposed hydrated sulfuric acid as the dominant non-ice component, and suggested that exogenous plasma implantation and endogenous geologic emplacement were both possible sources of sulfur. \cite{dal05} suggested that the reddish and dark terrains were a complex mixture of sulfate hydrates and other species. \cite{dal07} recognized that the NIMS spectra are better described by mixtures of salt and sulfuric acid hydrates, suggesting both to be present. \cite{mcc10} used a supervised spectral mixture analysis to claim five end members in one NIMS high resolution scene, and noted a distinction between the trailing hemisphere dark material and red material in linea. \cite{shi10} explored the association between NIMS spectra and terrain type in a NIMS high resolution scene, claiming a spatial distinction between sulfuric acid hydrate and salt hydrates, with a greater abundance of hydrated salts and not hydrated sulfuric acid in the low albedo plains. \cite{dal12} examined four NIMS high resolution slices and suggested three dominant spectroscopic components: water ice, hydrated sulfuric acid, and hydrated salts. Beyond NIMS, \cite{brown13} used hyperspectral observations with Keck II adaptive optics to identify a new spectral feature at 2.07~$\mu$m, and find that it is associated with the trailing hemisphere and exogenous bombardment.

Here we analyze the spatially resolved spectra from \cite{brown13}, with higher spectral resolution and greater geographic coverage than NIMS, and find evidence for three distinct compositional units at the $\sim$100~km scale. The endogenous and exogenous units are revealed by their geographical distributions; the exogenous component dominates the bullseye centered on the trailing hemisphere, and the endogenous component dominates leading hemisphere chaos units. A distinct composition associated with surface geology is not consistent with an ice shell passively altered by exogenous processes, but is consistent with active cycling of the ice shell. We compare the characteristic endogenous spectrum to laboratory spectra of plausible endogenous minerals, and find that it is not consistent with the hydrated sulfate salts previously favored. A definitive identification remains elusive, but chloride evaporite deposits are one possible alternative.

\begin{table*}
\begin{center}
\caption{Summary of Europa observations.\label{table1}}
\begin{tabular}{ccccccc}
\tableline\tableline
\\
Date (UT) & Time (UT) & Hemisphere & Limb & Band & Sub-Earth lon. & Sub-Earth lat.\\
\tableline
\\
2011 Sep 18 & 09:10 & Leading & west & K & 41 W &  3 N\\
2011 Sep 18 & 09:26 & Leading & east & K & 43 W &  3 N\\
2011 Sep 18 & 10:22 & Leading & west & H & 47 W &  3 N\\
2011 Sep 18 & 10:31 & Leading & east & H & 48 W &  3 N\\
\tableline
\\
2011 Sep 19 & 09:05 & anti-Jovian & west & K & 143 W & 3 N \\
2011 Sep 19 & 09:37 & anti-Jovian & east & K & 145 W & 3 N \\
2011 Sep 19 & 10:48 & anti-Jovian & west & H & 150 W & 3 N \\
2011 Sep 19 & 11:06 & anti-Jovian & east & H & 151 W & 3 N \\
\tableline
\\
2011 Sep 19 & 14:04 & anti-Jovian & west & K & 164 W & 3 N \\
2011 Sep 19 & 14:37 & anti-Jovian & east & K & 166 W & 3 N \\
2011 Sep 19 & 15:05 & anti-Jovian & west & H & 167 W & 3 N \\
2011 Sep 19 & 15:22 & anti-Jovian & east & H & 169 W & 3 N \\
\tableline
\\
2011 Sep 20 & 09:09 & Trailing & west & K & 244 W & 3 N \\
2011 Sep 20 & 09:43 & Trailing & east & K & 247 W & 3 N \\
2011 Sep 20 & 10:33 & Trailing & west & H & 250 W & 3 N \\
2011 Sep 20 & 10:51 & Trailing & east & H & 252 W & 3 N \\
\tableline\tableline
\\
\end{tabular}
\end{center}
\end{table*}

\section{Observations and data reduction}

Europa was observed over three nights in 2011 September 18-20 with the OSIRIS integral field spectrograph and adaptive optics system on the Keck II telescope \citep{lar03}. Observations were acquired at four orbital orientations within one Europa orbit, and give nearly complete latitude/longitude coverage. Spectra were obtained at moderate resolution (R~$\approx$~3800) in the infrared H and K bands (1.473-1.803 and 1.956-2.381 $\mu$m), at the 0.035$\arcsec$~pixel$^{-1}$ setting, and an angular resolution coarser than the detector resolution by a factor of $\sim$2. The disk of Europa was $\sim$1.03\arcsec on the sky, yielding an average surface resolution of $\sim$150~km per spatial element. The OSIRIS grid has a field of view 0.56\arcsec$\times$2.24\arcsec, so that the entire disk of Europa was captured with two exposures per spectral band. See table~\ref{table1} for the relevant orientations observed. For details regarding calibration, initial data reduction, and a complete journal of observations, see \cite{brown13}.

In addition to the calibration from \cite{brown13}, we make an extra correction to most accurately match spectra to specific locations on the surface. This task is slightly complicated by each hyperspectral full-disk consisting of four separate exposures, an east and west limb in the H and K bands (table~\ref{table1}). Due to Europa's rotation, exposures in H and K separated by $\sim$1 hour are offset in longitude by $\sim$6$^\circ$ or $\sim$1.6 OSIRIS pixels at the sub-observer point. To partially account for this, we use a lambertian surface to normalize each H and K observation to normal incidence. We then generate an artificial grid with the same resolution as OSIRIS, centered at the average sub-observer longitude of the four exposures. The full H and K spectrum of each location in the new grid is obtained by nearest interpolation of the original exposures. This results in a spatial error between H and K spectra of $<$1 OSIRIS pixel or $\lesssim$3.8$^\circ$ longitude at the sub-observer point.

\section{End member extraction}

In order to coherently map the surface composition of Europa, we first seek the smallest number of unique observed surface spectra which, when linearly combined, can be used to optimally model all of the spectra on the surface of Europa. While past efforts at such modeling have relied on matching laboratory-derived spectral libraries \citep{mcc99, car05, shi10, dal12}, or supervised selection of surface spectra \citep{mcc10}, we instead take a purely empirical approach in order to remove preconceived assumptions on potential surface compositions.

Mathematically, we define the $N$ end members as the spectra at $N$ different locations which, when used as the basis set in a linear least-squares fit of the separate spectrum at each spatial location, minimizes the sum of the square of the residuals across the entire surface of Europa. With this definition, our task is then to determine $N$, the appropriate number of end members, and what each end member spectrum is. There is an extensive literature on the subject of end member extraction and many possible approaches, and no single approach is preferred for all situations. We chose to develop our own unsupervised linear unmixing routine tailored for the OSIRIS data set.

We first filter out all spectra with obvious systematic noise, and those observed at more than 60$^\circ$ from the sub-solar point. This leaves 1,591 high quality spectra from the four observations combined. We bin these to a spectral resolution of 10~nm for computational feasibility. We input all spectra into k-means algorithm \citep{ped11} to find $N$ cluster center spectra. We then find the observed spectra most similar to the $N$ cluster centers, and set these as the initial guesses for candidate end members. This initial degree of similarity is determined by the spectral angle mapping (SAM) metric \citep{kru93}, where the angle $\theta$ between spectra is small for similar spectra and large for dissimilar spectra, and is defined: $\theta = cos^{-1}( \frac{ \bold{x \cdot y} }{ \bold{\| x \| \cdot \| y \| } } )$, where $\bold{x}$ and $\bold{y}$ are spectra. We take these $N$ spectra as candidate end members, and model each of the 1,591 observed spectra as a linear combination of the $N$ candidate spectra. For each of the 1,591 fits, the residual RMS is calculated and the RMS of each individual fit is summed to give the total RMS. 

\begin{figure}
\begin{center}
\includegraphics[scale=0.31]{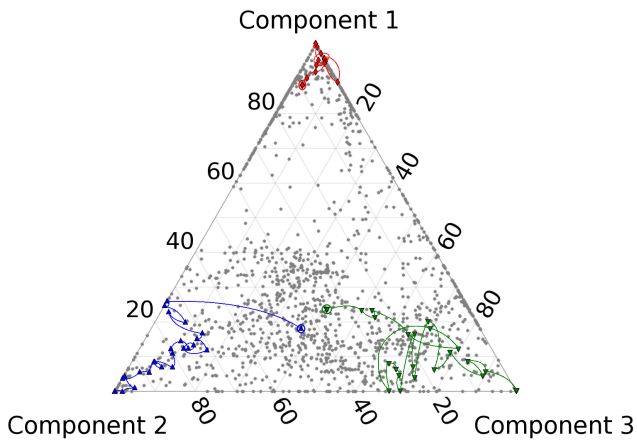}
\caption{Progression of the end member extraction routine for three end members. End members converge to the corners of the ternary plot (color symbols). Initial guesses are obtained from a standard k-means algorithm and SAM distance metric. Candidate end members are iteratively substituted with new candidates, and replaced if the new candidate lowers the total RMS. All observed spectra (gray symbols) are a combination of the end member components.\label{ternary}}
\end{center}
\end{figure}

To improve from the initial candidates, we hold all but one of the end member candidates fixed and substitute the remaining candidate with other observed spectra. As before, the substituted spectrum and fixed candidates are used to model all 1,591 observed spectra, and the RMS of each of the 1,591 fits is summed to give the total RMS. If the substituted spectrum yields a lower total RMS than the previous candidate, it becomes the new candidate. We iterate through all observed spectra to find the best new candidate. The process is repeated successively, alternating which end member is being substituted for. The process ends when the end members have converged, which occurs when a new candidate is not found for $N-1$ successive attempts. This process is illustrated for the $N=3$ case in the ternary plot in figure \ref{ternary}. The end member candidates take turns to converge to the best end members at the corners of the plot. 

\begin{figure*}
\begin{center}
\includegraphics[scale=0.6]{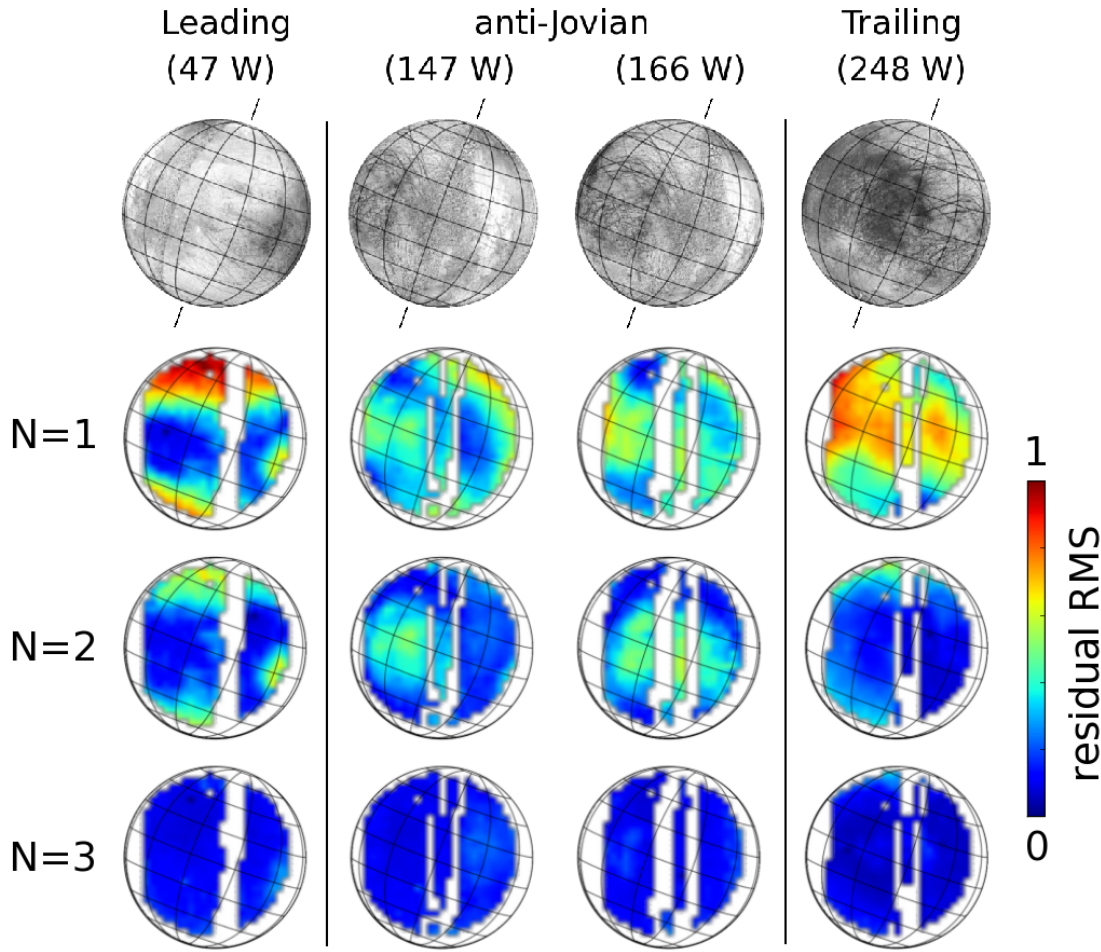}
\caption{Residuals of the end member extraction routine for 1, 2, and 3 end members from top to bottom. Grayscale globes show the orientation of Europa at the time of observation (USGS Map-a-planet: http://astrogeology.usgs.gov/tools/map). The four globes from left to right represent the four observations. For 1 and 2 end members, there remain large spatially coherent regions in RMS, indicating spectrally distinct regions not well described by 1 or 2 end members. The RMS level significantly drops for 3 end members, indicating 3 major, distinct compositions. Limbs are trimmed beyond 60$^{\circ}$ from the sub-solar point, which is offset from the sub-observer point by 8$^{\circ}$~W and 1$^{\circ}$~S. Gaps in the data result from stitching separate east/west and H/K observations into a complete hyperspectral cube.\label{RMS}}
\end{center}
\end{figure*}

To determine the most appropriate $N$, we use the simple approach of analyzing a range of $N$ and examining the successive residual maps. Once the residual deviation has dropped to the level of noise, spatial structure in the residual maps is lost and the number $N$ of end members used is optimal. We reach this threshold at $N=3$ (fig. \ref{RMS}). For $N=1$ and $2$, there is significant spatial structure in the residual maps, indicating compositional regions where the model is inadequate with one or two end members. The RMS level drops significantly from $N=1$ (total RMS = 42.7) to $N=2$ (total RMS = 23.6) and from $N=2$ to $N=3$ (total RMS = 11.4), removing nearly all spatial coherence in the residual map. $N=4$ (total RMS = 9.9) and $N=5$ (total RMS = 9.4) give only small improvements over $N=3$. Remaining structure in the $N=3$ residuals may indicate further end members, however any further end members would be much less significant, and it is unclear if the remaining structure is related to the physical surface or to systematic noise, so we conservatively choose $N=3$ as the optimal case. As a test of convergence, we repeat the routine for $N=3$ using random initial candidates instead of the cluster center candidates. In 50 tests with random initial guesses, the routine converged to only two results. Both results are qualitatively and quantitatively similar. The better result has slightly lower total RMS and was reached for 39 (78\%) of the random tests, and is the same result reached when using the initial guess from k-means + SAM.

\section{Compositional geography}

\begin{figure*}
\begin{center}
\includegraphics[scale=0.6]{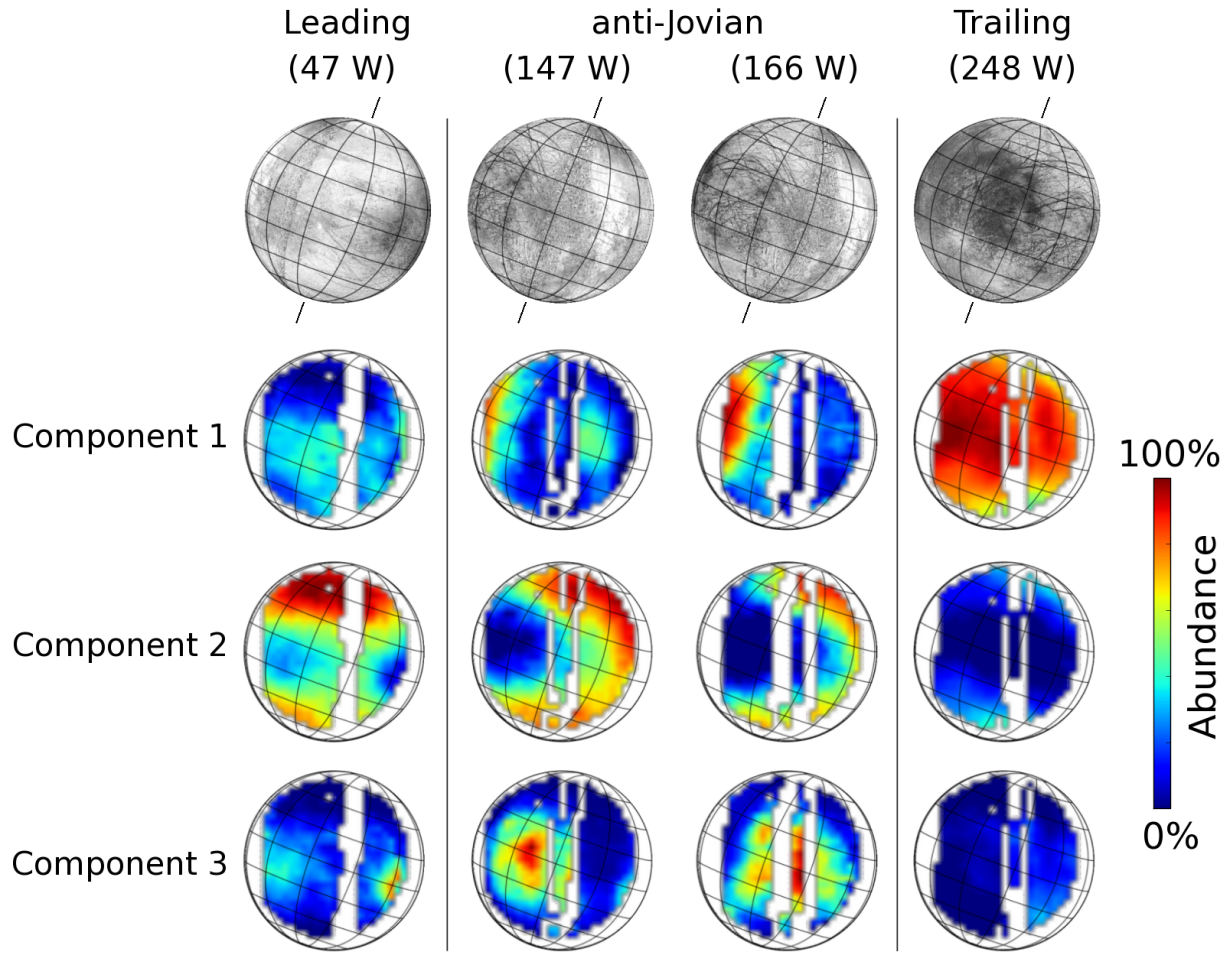}
\caption{Distribution of three end member components projected to the orientation at the time of observation. Grayscale globes show the orientation of Europa at the time of observation (USGS Map-a-planet: http://astrogeology.usgs.gov/tools/map). Component~1 dominates the trailing hemisphere, component~2 dominates the leading hemisphere upper latitudes, and component~3 dominates several regions across the leading hemisphere. The rotation of the component~3 regions between observations further confiRMS their existence. Limbs are trimmed beyond 60$^{\circ}$ from the sub-solar point, which is offset from the sub-observer point by 8$^{\circ}$~W and 1$^{\circ}$~S. Gaps in the data result from stitching separate east/west and H/K observations into a complete hyperspectral cube.\label{ortho}}
\end{center}
\end{figure*}

\begin{figure}
\begin{center}
\includegraphics[scale=0.6]{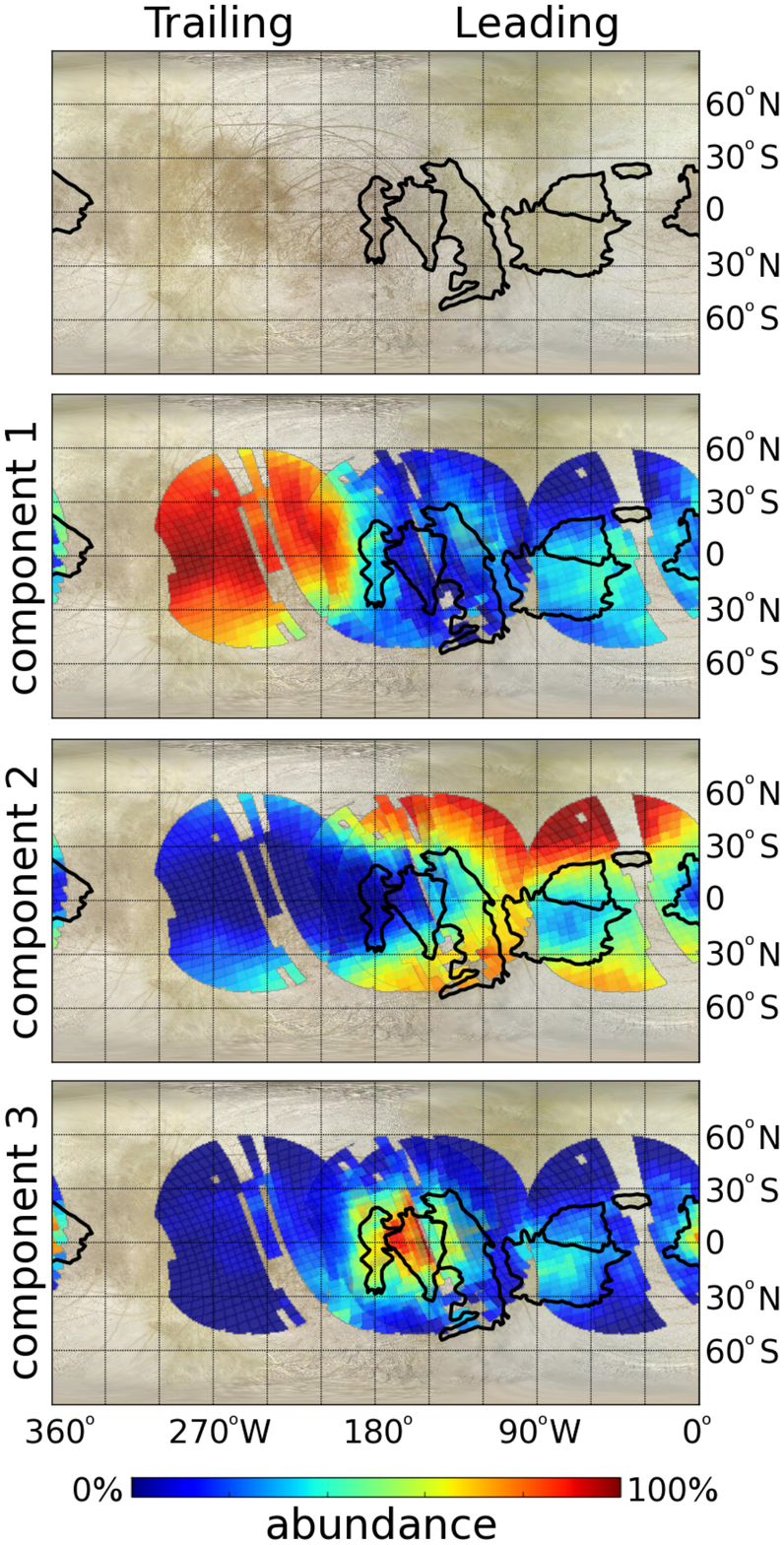}
\caption{Distribution of components~1, 2, and 3 mapped to a cylindrical projection. Overplotted in black are approximate locations of large leading hemisphere chaos regions from \cite{dog09}. Component~1 dominates the trailing hemisphere bullseye and component~2 dominates the leading hemisphere upper latitudes, in good agreement with known ``non-icy" and ``icy" Europa. Component~3 is associated with leading hemisphere chaos regions to within the angular resolution of the observations, suggesting an endogenous origin.\label{latlon}}
\end{center}
\end{figure}

With the approach outlined in the previous section, we find evidence for three distinct compositional units. Their geographic distributions are given by their fractional weights in the linear mixture model (fig. \ref{spec}). Their orthographic projections at the time of observation are shown in figure~\ref{ortho}, and their cylindrical projections are shown in figure~\ref{latlon}. Component~1 dominates the trailing hemisphere bullseye region, component~2 dominates the leading hemisphere upper latitudes, and component~3 dominates several regions with unique shapes across the leading hemisphere equatorial regions. The most distinct regions of component~3 clearly rotate between the two September~19 observations (fig. \ref{ortho}), further confirming their existence. The end member spectra are located in the middle of the trailing hemisphere (281~W, 4~S), the leading hemisphere northern mid-latitudes (77~W, 32~N), and within Western Powys Regio (155~W, 5~S) for components~1, 2, and 3, respectively. Though the end members occupy only one spectral pixel each, linear combinations of these three adequately reproduce all observed spectra across the surface, and yield compositional regions with impressive spatial coherence.

It is interesting that component~3 was not discovered previously in NIMS data. Though more geographically limited, NIMS low resolution data do reach the anti-Jovian regions dominated by component~3 \citep{car09}. However, previous work investigating compositional distributions at low spatial resolution \citep{mcc98, mcc99, car05, gru07, brown13} focus on ice vs. hydrate fractional abundance, and do not attempt to distinguish between different hydrate species. The distinction between different hydrated minerals has been explored in NIMS high resolution data \citep{dal05, dal12, dal07, mcc10, shi10}, which lacks the necessary surface coverage distinguish global trends. In fact, the anti-Jovian regions dominated by component~3 are distinguishable in NIMS low resolution data, but were not previously recognized as a distinct composition. Western Powys Regio particularly stands out in several water-ice abundance maps \citep{gru07, mcc99, car05}, but was not recognized as distinct from the dominant non-ice component centered on the trailing hemisphere bullseye. Western Powys and the adjacent chaos region at 180 W longitude are most noticeable in the quality of fit map in plate~2 of \cite{mcc99}, where the poorer quality of fit is clearly associated with these regions, indicating a distinct composition. These regions are also seen in \cite{han13}, which maps the sulfuric acid abundance and average hydrate abundance to NIMS low resolution data; the regions corresponding to component~3 stand out in average hydrate abundance but not in sulfuric acid abundance. Thus NIMS data are consistent with our results.

\section{Discussion: identifying composition}

\begin{figure}
\includegraphics[scale=0.33]{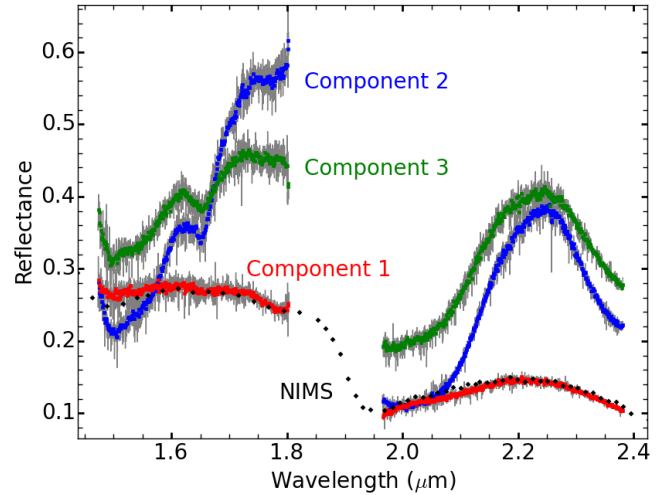}
\caption{OSIRIS spectra of Europa's three dominant end members at the $\sim$100 km scale. Gray lines show the true spectral resolution; colored spectra are binned to a resolution of 1~nm. The NIMS Europa hydrate spectrum \citep{car99} is overplotted in black and scaled for comparison. Note the similarity of NIMS to component~1, though at a much coarser resolution.\label{spec}}
\end{figure}

The observed spectra of the three compositions are shown in fig.~\ref{spec}. We model each hyperspectral pixel as a linear combination of these three end member spectra, and find that the entire surface at the $\sim$100~km scale is sufficiently described by a mixture of these three components. Though all three components share the same major spectral features due to water of hydration, they are clearly distinct from each other. Component~1 resembles a darkened and broadened ice spectrum. The features of component~2 are the least distorted and have the strongest band depths, indicating that this component has the greatest fraction of pure water-ice. Component~3 resembles a flattened water-ice spectrum with slightly broadened and shifted features. Distorted water-ice spectra such as this suggest hydrated species \citep{mcc98, mcc99, mcc02}, though the lack of additional features makes a unique identification difficult.  

The geographic distribution of each component is a strong indication of its identity. The distribution of component~1 is globally symmetric, centered on the trailing hemisphere equator. This is consistent with the known patterns of enhanced irradiation and bombardment from the Jovian magnetosphere \citep{car99, car02, par01, cas13, dal13b}. This strongly suggests that component~1 is exogenous, or initially endogenous but heavily altered by exogenous processes. Component~2 dominates the leading hemisphere polar regions, consistent with the known distribution of pure water ice or deposited water ice frost \citep{gru07, brown13}. Unlike components~1 and 2, the distribution of component~3 is not globally symmetric, nor is it consistent with regions known to be compositionally distinct. The shapes of the regions dominated by component~3 are revealing; they are markedly similar to geologic units of chaos as mapped by \cite{dog09}, to within the angular resolution of our observations. Though the spectral pixel identified as component~3 is located within Western Powys Regio, it also dominates the abundance in other regions; Tara Regio chaos in particular is reproduced remarkably well (figs. \ref{ortho}, \ref{latlon}, $\sim$85~W). Chaos regions are believed to represent areas of recent interaction with the subsurface, and a variety of mechanisms have been proposed including widescale diapiric upwelling \citep{pap98}, melt-through \citep{gre99}, and lens-collapse \citep{sch11}. Thus, the association of component~3 with chaos suggests an initially endogenous composition, less altered by exogenous processes than the trailing hemisphere. The trailing hemisphere is also covered with large chaos regions, in which we see no evidence for component~3. Our interpretation is that component~3 contains the greatest fraction of salts derived from the subsurface, possibly a brine or evaporite deposit. The lack of component~3 on the trailing hemisphere is due to enhanced exogenous alteration, where the surface has been coated with a layer of irradiation products, primarily sulfuric acid hydrate \citep{car99}. An endogenous composition likely exists at greater depth in trailing hemisphere chaos regions, where it is shielded from exogenous processes and NIR remote sensing.

\begin{figure}
\begin{center}
\includegraphics[scale=0.38]{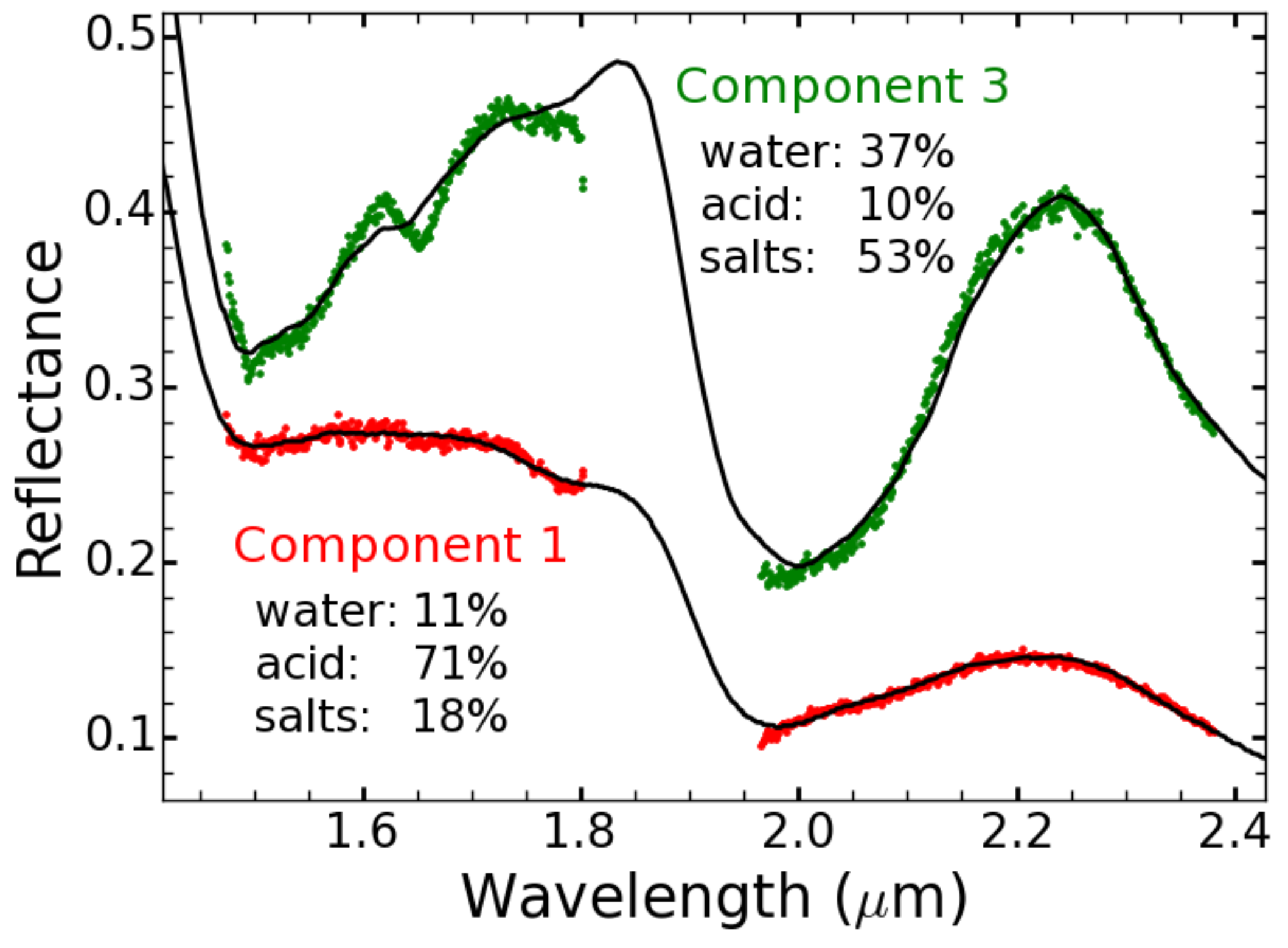}
\caption{Linear mixture fits to the non-ice components, following the method of \cite{dal07}. At moderate spectral resolution, this library describes component~1 well and component~3 poorly. The result for component 1 is 71\% H$_2$SO$_4 \cdot$nH$_2$O, 11\% bloedite, 7\% 250~$\mu$m H$_2$O ice, 4\% mirabilite, 4\% 1000~$\mu$m H$_2$O ice, 3\% hexahydrite, and $<$1\% MgSO$_4$ brine and H$_2$O ice at 50, 75, 100~$\mu$m. The result for component~3 is 42\% bloedite, 37\% 50~$\mu$m H$_2$O ice, 11\% mirabilite, 10\% H$_2$SO$_4\cdot$nH$_2$O, and $<$1\% MgSO$_4$ brine, hexahydrite, and H$_2$O ice at 75, 100, 250, and 1000~$\mu$m.\label{fits}}
\end{center}
\end{figure}

It was suggested from NIMS spectra that hydrated salts are abundant on Europa's surface, particularly Na- and Mg-sulfates \citep{mcc98}. Laboratory spectra of these salts have been used extensively to model high spatial resolution NIMS data \citep{dal05, dal12, dal07, shi10}. Following these analyses, we perform similar linear mixture modeling to the OSIRIS data. The results are shown in figure~\ref{fits}. In the fits shown, we include H$_2$SO$_4$ hydrate, mirabilite (Na$_2$SO$_4\cdot$10H$_2$O), hexahydrite (MgSO$_4\cdot$6H$_2$O), bloedite (Na$_2$Mg(SO$_4$)$_2\cdot$4H$_2$O), MgSO$_4$ brine, and water ice at grain sizes of 50, 75, 100, 250, and 1000 $\mu$m. The fit to component~1 is generally satisfying. This method finds that component~1 is dominated by H$_2$SO$_4$ with lesser amounts of the modeled salts and water ice, consistent with an exogenous interpretation. Furthermore, this method finds that component~3 is slightly dominated by hydrated salts, with a significant but lesser amount of fine-grained water ice and little H$_2$SO$_4$ hydrate, consistent with an endogenous interpretation for component~3. 

\begin{figure}
\begin{center}
\includegraphics[scale=0.41]{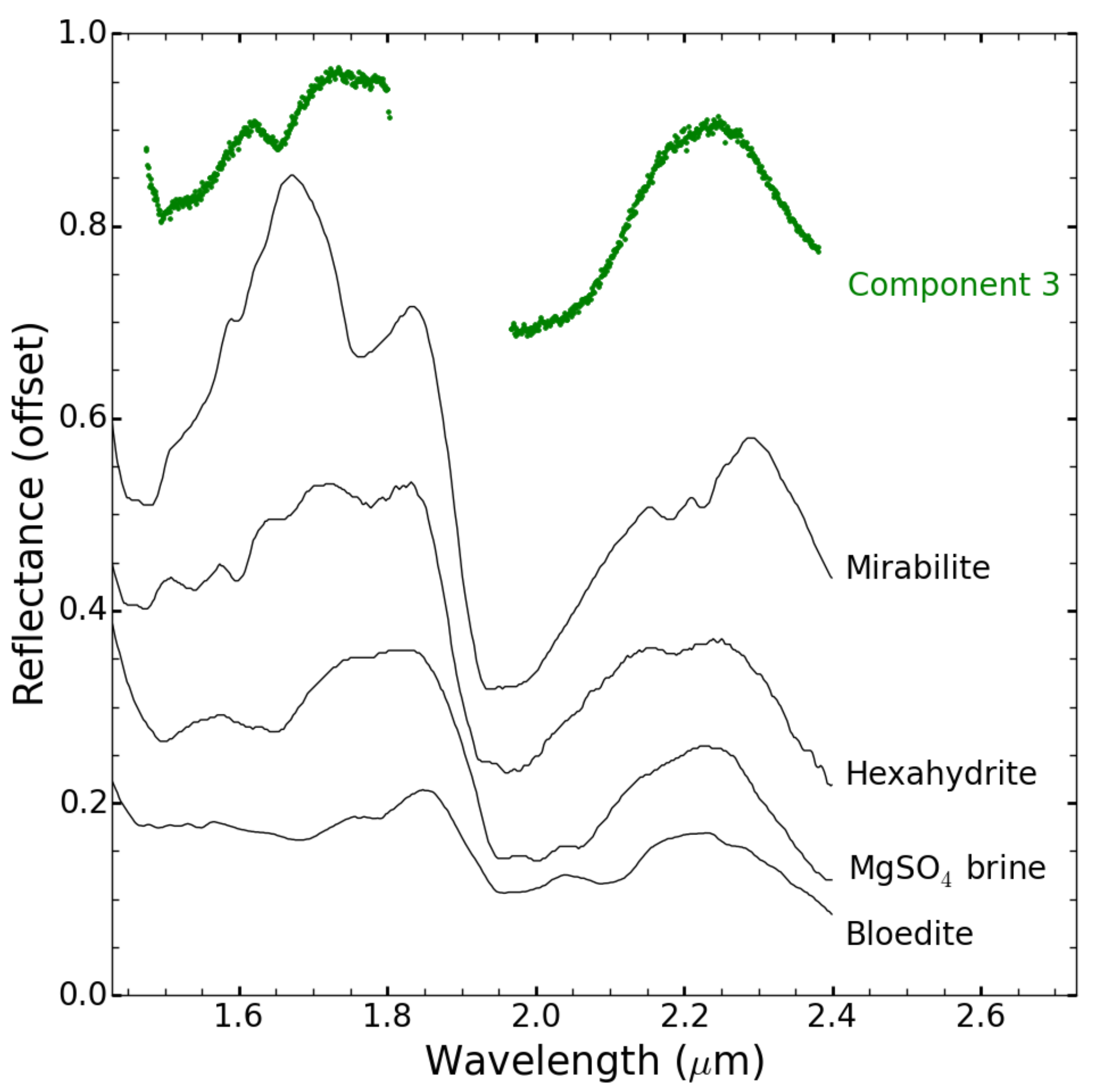}
\caption{Spectrum of component~3 compared to spectra of various hydrated sulfate salts \citep{dal05} traditionally thought to be abundant on Europa's surface. Note their spectral dissimilarity to the OSIRIS spectrum of component~3, representing potentially the saltiest composition on Europa. At sufficiently fine spectral resolution, hydrated sulfate minerals possess many distinct features not seen in the observed spectra, constraining their surface abundances. Spectra are offset for clarity by 0.5, 0.13, 0.07, -0.18, and 0.0 from top to bottom.\label{nonice}}
\end{center}
\end{figure}

Though this result is enticing, the quality of the spectral fit is unacceptably poor (fig. \ref{fits}). Although a linear mixture of these species is able to match the coarse continuum level, it is not able to fit the shape and depth of distinct features in the H band, or the distorted shape of the 2.2~$\mu$m peak in the K band. This remains true even when we relax the constraints of the linear mixture model: in the fits shown, we include a dark neutral substance with a spectral slope, and do not restrict the sum of the components to match 100\%. Figure~\ref{nonice} compares the observed spectra to the laboratory spectra. Conversely, figure~\ref{nonice} shows significant differences between component~3 and the hydrated sulfate salts considered. Each of these salts possesses distinct features that should be seen in the observed spectra, if they were present in abundance. Though the results of linear mixtures with these salts are consistent with an endogenous interpretation, the individual sulfate salts considered cannot be present in abundance. 

\begin{figure}
\begin{center}
\includegraphics[scale=0.37]{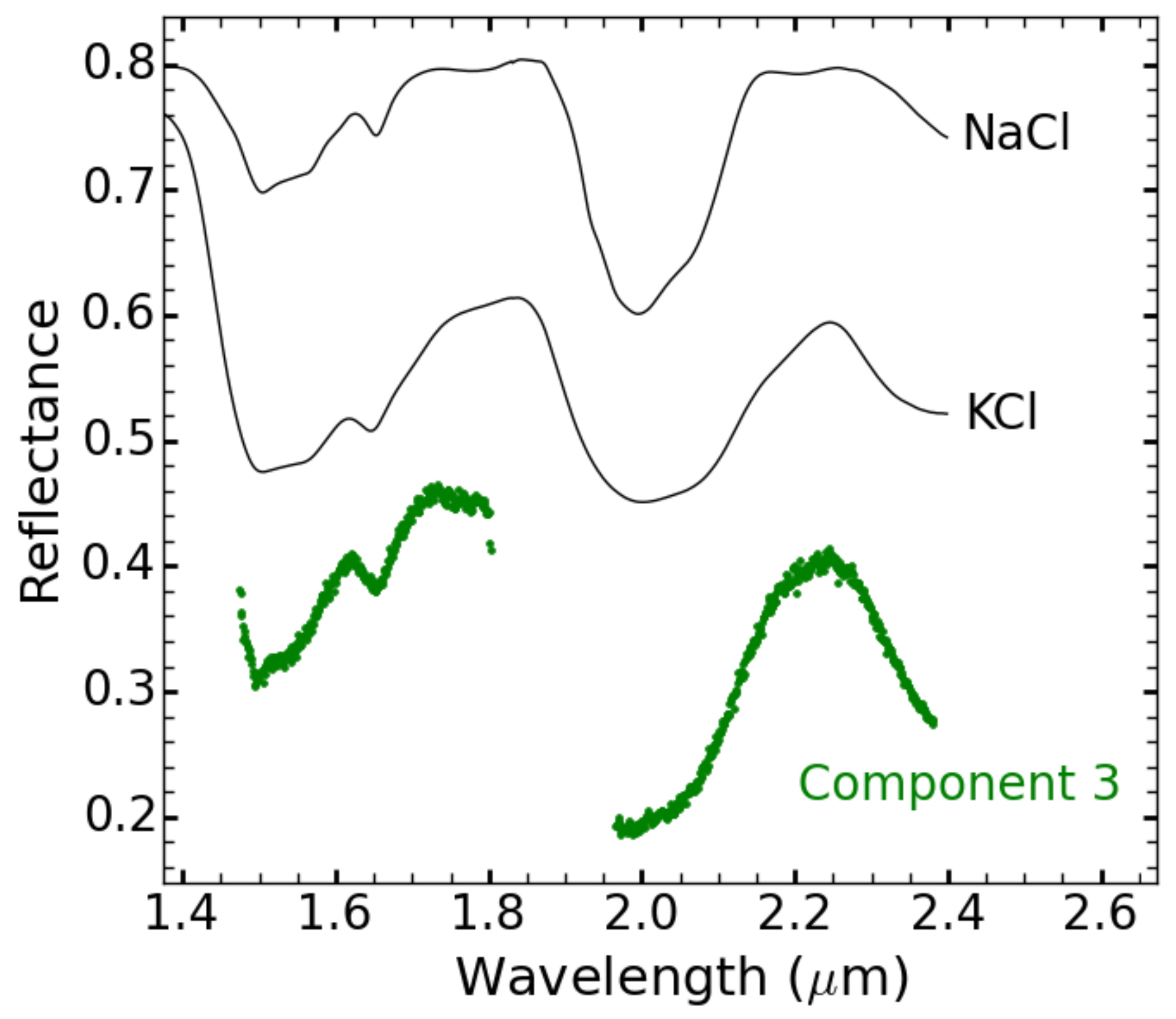}
\caption{Spectrum of component~3 compared to alternative salt compositions. Anhydrous NaCl and KCl \citep{brown13} are similarly smooth. Importantly, they lack the additional features characteristic of hydrated sulfate minerals that are not seen in the observed spectra.\label{salt}}
\end{center}
\end{figure}

At the observed wavelengths, the defining spectral characteristics of component~3 are the distortion of water-of-hydration features and the lack of additional, non-water-ice features. Thus the species making up component~3 must share these characteristics. One possible alternative composition is anhydrous chloride salts such as NaCl, KCl (fig.~\ref{salt}). Note that the term anhydrous is relative, e.g. baking at 400$^\circ$C for days is not sufficient to remove all of the bound water, so residual features persist \citep{han14}. In the case of NaCl, the anhydrous phase is more spectrally similar to the observed spectra than the brine phase \citep[Figures 1 and 4 in][]{han14}. However, the chlorate and perchlorate salts from \cite{han14} possess distinct spectral features not seen in the observed spectra. Anhydrous salts could indeed be present on Europa's surface; evaporite deposits are speculated to result from exposure of a brine to Europa's surface conditions through the processes of sublimation, freezing, and flash evaporation \citep{mcc98}. A search through the literature of published spectra at conditions relevant to Europa and spectral resolution comparable to OSIRIS revealed NaClO and NaOH \citep{brown13}, which are conceivable radiolytic products of a NaCl-H$_2$O substrate, and NaHCO$_3$ brine \citep{dal05}, which is a plausible endogenous species. Other carbonate salts are also plausible, but must be investigated at appropriate conditions and moderate spectral resolution.

\section{Discussion: revisiting the sulfate hypothesis}

The traditional view that Europa's endogenous non-ice component is dominated by Mg-sulfates traces back at least as far as \cite{kar91}, which found the system MgSO$_4$-Na$_2$SO$_4$-H$_2$O to be the favored product of aqueously differentiated carbonaceous chondrites in conditions expected for icy satellites. This served as a working hypothesis for early NIMS investigations such as \cite{mcc98}, which found an expected spectral agreement between NIMS and sulfate salts. This story was further supported by chondrite-leaching experiments \citep{fan01} and subsequent geochemical modeling \citep{zol01}. Later NIMS investigations showed that the spectra are better explained by mixtures of sulfuric acid and hydrated salts, but still focus solely on sulfate salts \citep{dal07, shi10, dal12, dal13a}. These results made for a satisfying, consistent story. However, there are many caveats to this story, and a growing body of work is challenging the traditional view. 

The spectroscopic argument set in motion by early NIMS investigations relied on an even smaller spectral library than exists today, among which hydrated sulfate salt minerals were one of the few items. Furthermore, NIMS investigations face the challenge of low spectral resolution; the spectral similarity amongst hydrated species makes the ability to spectrally distinguish between them difficult, and this difficulty is exacerbated at low spectral resolution, where small features are smoothed out and the subtle but significant shape of the continuum is less apparent. OSIRIS observations show that the NIR moderate resolution spectra of the entire surface resembles water-ice, precluding the possibility for species with distinct non-ice spectral features. The only exception to this is the 2.07~$\mu$m feature attributed to epsomite \citep{brown13}. Ironically, the identification of epsomite challenges the case for endogenous sulfates, due to the geographical association of epsomite with exogenous and not endogenous processes. The OSIRIS spectra are in good agreement with similar observations from \cite{spe06}, which resolve ice and non-ice units and find equally smooth, featureless spectra. \cite{spe06} also conclude that higher resolution spectra are inconsistent with the crystalline hydrates considered.

Geochemical models predicting surface composition also claim a large degree of uncertainty. \cite{kar91} notes that sufficiently alkaline conditions during aqueous alteration would stabilize Mg and Fe in Mg/Fe-hydroxides, preventing  significant amounts of Mg in the solution, or that sufficiently acidic conditions could favor more abundant carbonates in solution. \cite{kar00} explores a wide variety of alternative surface compositions, and for example notes that the hydrothermal processes causing a NaCl ocean on Earth could also occur on Europa under the right conditions. \cite{mck03} raise many objections to the models favoring a MgSO$_4$-dominated system, including a possibly incorrect initial composition, the irrelevant conditions of chondrite leaching experiments of \cite{fan01}, and the difficulty of forming sulfate in reducing conditions. However, though surface and ocean composition are certainly related, the exact relationship is uncertain; an alternative surface composition suggests but does not necessarily require an alternative ocean composition. For example, \cite{zol01} point out that freezing leads to brines of different composition than the ocean, and that endogenous surface salts may only be the end member of a compositional stratification throughout the ice shell.

Europa's known atmospheric composition further challenges the traditional view. The tenuous atmosphere is produced by sputtering of surface material \citep{joh98}; thus surface and atmosphere compostion are intimately related. The presence of Na \citep{brown96} and K \citep{brown01} in the sputtered atmosphere is established, and their relative abundances provide a compelling argument for an endogenous source \citep{brown01}. Furthermore, the non-detection of atmospheric Mg \citep{hor13} constrains its surface abundance and challenges the traditional view of an endogenous surface component dominated by the Mg cation.

Enceladus may represent an analogous environment to Europa, and Enceladus' plume composition also suggests an alternative salty component. Na, K, Cl, and CO$_3$ have all been detected, whereas Mg and SO$_4$ have not \citep{post09, post11}. Using a geochemical model based on the empirical evidence, \cite{gle15} favor an Enceladus ocean composition of Na-Cl-CO$_3$.  

Furthermore, recent work by \cite{kph15} shows that irradiated alkali halides such as NaCl can describe the yellow to brown colors and coarse continuum level of Europa's visible spectrum. The formation of similar color centers in sulfate salts is less certain but many sulfates bleach to white rapidly after irradiation or stay colorless in the visible even after irradiation \citep[][Hand, K.~P. et al., in preparation]{nas77}. Previously, the darker colors seen by \textit{Galileo} Solid State Imager (SSI) were attributed to a mixture with an unknown associated species that is featureless in the infrared \citep{dal05, mcc10}. 

This work further supports that the traditional view of Europa's endogenous surface composition must be reconsidered. We find evidence that Europa's surface consists of three distinct compositional units at the $\sim$100~km scale. We identify the endogenous composition based on association with geologic chaos units, and show that its spectrum is inconsistent with hydrated sulfate minerals at moderate resolution. 

The implications of alternative salts for habitability are intriguing but difficult to interpret, even with a complete knowledge of surface composition and subsurface exchange. We can speculate that the dominant ocean salt has implications for salinity, which may be a limiting factor for the origin of life and the ability for life to thrive. Though the salinity of the Europan ocean is only marginally constrained, an ocean dominated by Na$^+$ and Cl$^-$ could imply a lower salinity compared to Mg$^{2+}$ and SO$_4^{2-}$, by a factor of $\sim$3 for the upper limit of measured ocean conductivity \citep{kph07}. We can also speculate that biological processes, if present, may manifest themselves in the endogenous surface composition. For example, a Europan biosphere supported by sulfate reduction \citep{zol03} could result in an ocean depleted in sulfate, if the supply of sulfate is the limiting factor for biomass production. This scenario is likely inconsistent with a surface abundant in endogenous sulfate. Similarly, a Europan biosphere supported by methanogenesis could yield a relatively reduced methane-rich ocean or a relatively oxidized sulfate- and bicarbonate-rich ocean \citep{mccol99}, leading to an endogenous surface enhaced in the respective species. However, abioligical processes are likely able to explain surface composition as well, and such interpretations should be favored while definitive evidence for biological activity is lacking.

\section{Conclusions}

We find evidence for three distinct compositions across Europa's surface at the $\sim$100 km scale, from hyperspectral observations with Keck II OSIRIS and adaptive optics. The first component dominates the trailing hemisphere bullseye and the second component dominates the leading hemisphere upper latitudes, consistent with regions previously found to be dominated by irradiation products and water ice, respectively. The third component is geographically associated with large geologic units of chaos, suggesting an endogenous identity. This is the first time that the endogenous hydrate species has been mapped at a global scale. 

Europa's moderate resolution surface spectra in the NIR are ubiquitously smooth, distorted water-ice spectra. The spectrum of component~3 is not consistent with linear mixtures of the current spectral library. In particular, the hydrated sulfate minerals previously favored possess distinct spectral features that are not present in the spectrum of component~3, and thus cannot be abundant at large scale. One alternative composition is chloride evaporite deposits, possibly indicating an ocean solute composition dominated by the Na$^+$ and Cl$^-$ ions.

The traditional view that Europa's non-ice endogenous surface is dominated by hydrated Mg-sulfates must be reconsidered. Higher resolution surface spectra at greater geographic coverage, as well as insight from new laboratory irradiation experiments, atmospheric composition, and analogy to Enceladus all point toward a new understanding of Europa's endogenous hydrate. 

\acknowledgments

This research was supported by Grant 1313461 from the National Science Foundation. K. P. H. acknowledges support from the Jet Propulsion Laboratory, California Institute of Technology, under a contract with the National Aeronautics and Space Administration and funded in part through the internal Research and Technology Development program. The data presented herein were obtained at the W. M. Keck Observatory, which is operated as a scientific partnership among the California Institute of Technology, the University of California, and the National Aeronautics and Space Administration. The Observatory was made possible by the generous financial support of the W. M. Keck Foundation. The authors wish to recognize and acknowledge the very significant cultural role and reverence that the summit of Mauna Kea has always had within the indigenous Hawaiian community. We are most fortunate to have the opportunity to conduct observations from this mountain.

\end{document}